# Rating the Crisis of Online Public Opinion Using a Multi-Level Index System

Fanqi Meng
School of Computer Science, Northeast Electric Power University, China
Guangdong Atv Academy for Performing Arts, Guangdong 523710, China
mengfanqi@neepu.edu.cn

Xixi Xiao
School of Computer Science, Northeast Electric Power University, China
xiaoxixi@neepu.edu.cn

Jingdong Wang
School of Computer Science, Northeast Electric Power University, China
wangjingdong@neepu.edu.cn

**Abstract:** *Online public opinion usually spreads rapidly and widely, thus a small incident probably evolves into a large social crisis in a very short time, and results in a heavy loss in credit or economic aspects. We propose a method to rate the crisis of online public opinion based on a multi-level index system to evaluate the impact of events objectively. Firstly, the dissemination mechanism of online public opinion is explained from the perspective of information ecology. According to the mechanism, some evaluation indexes are selected through correlation analysis and principal component analysis. Then, a classification model of text emotion is created via the training by deep learning to achieve the accurate quantification of the emotional indexes in the index system. Finally, based on the multi-level evaluation index system and grey correlation analysis, we propose a method to rate the crisis of online public opinion. The experiment with the real-time incident show that this method can objectively evaluate the emotional tendency of Internet users and rate the crisis in different dissemination stages of online public opinion. It is helpful to realizing the crisis warning of online public opinion and timely blocking the further spread of the crisis.*

## 1. Introduction

Online public opinion represents netizens' emotions, attitudes and opinions. Take the "Xi'an female Benz owner safeguards rights" as an example, the customer disputes in the car store spread instantly on the Internet. Meanwhile, various news and rumors also continuously spread, which caused a terrible impact on the car owner, Benz car dealers, Mercedes-Benz Company and even the whole car sales industry and the regulatory authorities in the end. In order to reduce this kind of harm as much as possible, it is necessary to monitor online public opinion and assess the level of crisis, so as to carry out crisis early warning in time and intervene in it at the right time [7]. Therefore, establishing a complete and objective evaluation index system has become a priority issue that must be solved nowadays.

At present, the research on online public opinion mainly focuses on the social network communication model [19], online public opinion monitoring and early warning [27], online public opinion control and guidance [20], etc., Although scholars have constructed various index systems for monitoring and evaluating online public opinion from different perspectives, there are relatively few index systems built in combination with the characteristics of online public opinion, and most of them only use a few of simple statistical index, such as the number of readings, the number of comments, likes, etc., [13, 23]. Not only the evaluation is not comprehensive, but there is also information overlap between the indicators, making it difficult to achieve accurate evaluation. Individual studies have considered more complex index [3, 4, 5], however, these indexes, are often vaguely defined and difficult to quantify.

Aiming at the above problems, we studied the mechanism of online public opinion dissemination according to the theory of information ecology. On the basis of this, we combined Correlation Analysis (CA) and Principal Component Analysis (PCA) to select evaluation indexes, and then constructed a multi-level online public opinion evaluation index system. We also employed deep learning techniques to train and establish a sentiment classification model of online public opinion text for the accurate quantification of sentiment indexes in the index system. Finally, we completed the crisis level evaluation of the online public opinion by using the index system and grey correlation analysis.

The novelty of this paper lies in the following two aspects. First, we introduce the sentiment tendency of Internet users into our multi-level evaluation index system, and propose a sentiment classification model of online public opinion text to quantify the sentiment tendency, so the evaluation of online public opinion can more accurate. Second, we divide the online public opinion crisis into 4 levels and propose a method to create the 4-level crisis model by adopting Delphi, so the warning of online public opinion crisis can be more timely.

The rest of this paper is organized as follows: Section 2 reviews related work. Section 3 explains how to build the multi-level evaluation index system, including index selection, data collection, index quantification, and crisis level division, etc., Section 4 combines with examples to calculate the changes of online public opinion on a certain moment, so as to confirm the practicability of the index system. Section 5 discusses the limitations of current research. Section 6 summarizes the paper and gives possible future research directions.

## 2. Related Work

### 2.1. Index System Design

Online public opinion also known as network opinion or online public opinion, etc. In its early research, Dai and Fei [6] constructed a network public opinion security evaluation index system based on the communication theme, which analyze the target theme from security perspective. Lan [9] constructed a three-dimensional evaluation index system of emergency online public opinion security based on the action mechanism and evolution law of emergency online public opinion. The three dimensions are netizens' response, information characteristics of emergencies and the diffusion of emergencies. Zhang [26] built an emergency network public opinion risk evaluation index system with a total of 21 index at three levels. Li *et al*. [10]. Proposed an index system for quantitative evaluation of the influence degree of network public opinion, and a quantitative calculation method for the influence degree of network public opinion based on the index system. Zhao *et al*. [27] evaluated the public opinion of China's Weibo by using the public opinion index system, which mainly includes four aspects: information source index, regional index, topic index and industry index. Lian *et al*. [11] established a strictly deduced topology model based on the complex network theory to describe the evolution of network public opinion. For example, the index system often lacks the analysis of the sentiment tendency of Internet users, and the evaluation is not comprehensive enough; some evaluation indexes are vaguely defined and difficult to measure; the rationality of the index system lacks experimental verification, etc.

### 2.2. Early Warning of Public Opinion Crisis

With the rise of machine learning, some scholars began to study the quantification of public opinion index system and crisis warning. Xing [21] constructed a negative network public opinion monitoring index system based on information entropy in the new media environment, and gave the influence level of public opinion to realize the monitoring of public opinion. Yin *et al*. [24] established a network public opinion event early warning model by introducing the Bayesian network modeling method. He built the network nodes by discretizing the content of the index system and learned the parameters. Finally, he used the obtained Bayesian network to predict the probability of public opinion being in a high, medium and low trend. Kim and Laskowski [8] established a public opinion early analysis and Early Warning System (EWS) to monitor and analyze public health through the system as well. Therefore, the timeliness and accuracy of its early warning are difficult to be guaranteed.

## 3. Evaluation of Public Opinion Crisis

### 3.1. General Idea

The overall idea to realize the evaluation of the crisis rating of online public opinion is shown in Figure 1.

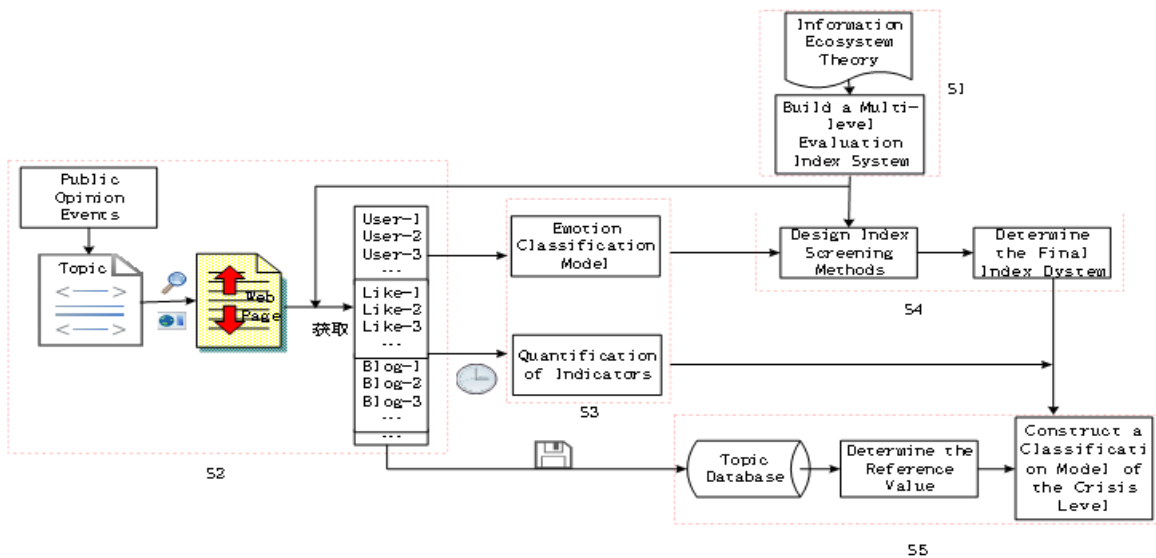

Figure 1. The overall idea to realize the evaluation of the crisis level of online public opinion.

- *Step* 1 (S1): build a multi-level evaluation index system of online public opinion including sentiment indexes (see section 3.2).
- *Step* 2 (S2): design a data collector to automatically collect online public opinion data (see section 3.3).
- *Step* 3 (S3): build an emotion classification model to quantify the emotion index based on the collected online public opinion text, (see section 3.4).

Considering that there may be problems such as information overlap in the initially established index system

- *Step* 4 (S4): determine the final index system by deleting some non-critical indexes (see section 3.5).
- *Step* 5 (S5): on the basis of the above steps, construct a classification model of the crisis level of online public opinion to achieve the objective evaluation and early warning ( see section 3.6).

### 3.2. Initial Construction of Indicator System

#### 3.2.1. Analysis of the Evolution

In order to classify the level of crisis properly, it is necessary to understand the stage of online public opinion evolution. The development of online public opinion usually goes through several stages, and the degree of crisis in different stages is not the same. Therefore, in order to build an objective and comprehensive index system, it is necessary to understand the evolution of online public opinion. The theory of information ecology believes that the evolution of online public opinion is nonlinear, presenting a spiral upward trend [12]. The information sending process and receiving process go back and forth to promote the continuous flow and dissemination of information. Figure 2 shows the general situation of online public opinion development, where the abscissa represents time, the online public opinion cycle is divided into five different periods, and the ordinate represents the level of evolution. Curve OM is a spiral curve trend diagram of online public opinion evolution.

In the Figure 2, A, B, C, D, and E represent the state of online public opinion evolution in different periods. Among them, A represents the period when the online public opinion has just appeared, it is called the budding period. B represents the online public opinion information began to spread around under the intervention of some online promoters and media, it is called the growth period in this paper. With the intervention of the government and the inflow of more real information, fake news is excluded, and Internet users with similar or identical views begin to form a small-scale combination, this is C, it is called the outbreak period. D called the fusion period, these small-scale combinations began to merge each other in the case of a high degree of mutual understanding with the continuous evolution of the online public opinion. In period E, the government decomposes these mainstream online public opinion information for finding the resolution. If the government and Internet users can meet each other's demands of interest the online public opinion of the emergency will calm down and the online public opinion information. And we call this recession. To avoid the outbreak of online public opinion events, it is necessary to discover the potential crisis in phase B in time, avoid phase C and enter phase e of intervention. Details can be seen in section F.

#### 3.2.2. Analysis of Factors Affecting the Spread of Online Public Opinion

In order to build an objective and comprehensive index system, we need to understand the characteristic factors of the spread of online public opinion. According to the information ecosystem theory, it can be found that the characteristic of online public opinion is composed of information factors, information technology factors, information people factors, and information environment factors. Online public opinion information refers to various carriers that reflect the state of public opinion and its movement. The corresponding relationship between these factors and each part of the online public opinion ecosystem is shown in Figure 3.

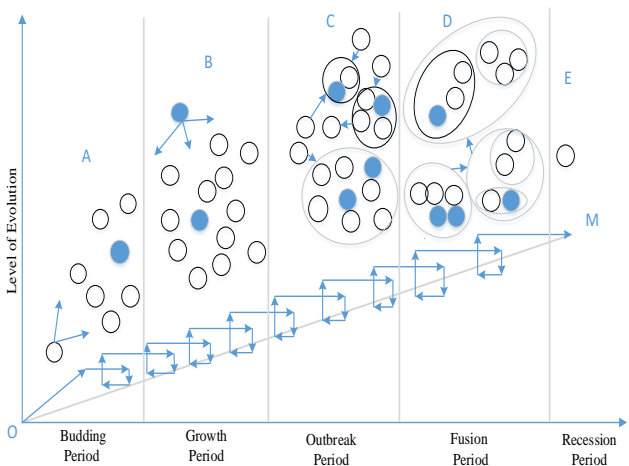

Figure 2. Evolution of online public opinion [12].

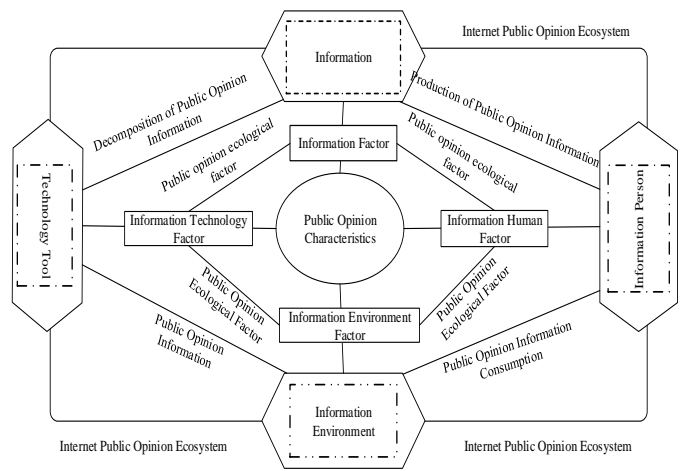

Figure 3. Influencing factors and characteristics of online public opinion [13].

### 3.2.3. Build an Index System Architecture

A preliminary multi-level evaluation index framework for online public opinion was constructed based on the analysis of online public opinion evolution situation and factors affecting online public opinion dissemination, see Table 1. It should be noted that all the numbers in the table is related to the same public opinion event.

$AvgFollower = \frac{\sum_{i=1}^{n} x_i - \max_{i=1...n}\{x_i\} - \min_{i=1...n}\{x_i\}}{n-2}$ is the average number of the followers or fans the bloggers who published original blogs have. Where $n$ is the total number of the bloggers, $x_i$ is the number of the followers or fans of the $i$-th blogger.

$AvgAttention = \frac{\sum_{i=1}^{n} x_i - \max_{i=1...n}\{x_i\} - \min_{i=1...n}\{x_i\}}{n-2}$ is the average number of the attentions the bloggers who published original blogs paid on other bloggers. Similarly, $n$ is the total number of the bloggers, $x_i$ is the number of attentions of the $i$-th blogger.

$AvgGrade = \frac{\sum_{i=1}^{n} x_i - \max_{i=1...n}\{x_i\} - \min_{i=1...n}\{x_i\}}{n-2}$ is the average rank of the original bloggers. Where $x_i$ is the grade of the $i$-th blogger.

V number is the number of certified VIP bloggers who originally published (exclude forwarding) a blog about the public opinion event.

$AvgHisVol = \frac{\sum_{i=1}^{n} x_i - \max_{i=1...n}\{x_i\} - \min_{i=1...n}\{x_i\}}{n-2}$ is the average number of blogs published by all the bloggers since their registration. Where $x_i$ is the historical blog volume of the $i$-th blogger.

Table 1. Initially constructed multi-level online public opinion evaluation index system.

| First-level index | Secondary-level index | Third-level index |
|---|---|---|
| Information Person | Internet User Importance | Average followers (*AvgFollower*) |
| | | Average attentions (*AvgAttention*) |
| | | Average grade (*AvgGrade*) |
| | | V number |
| | | Average historical blog volume (*AvgHisVol*) |
| | Internet User Participation | Total number of likes of online public opinion events |
| | | Total number of comments on online public opinion events |
| | | Total number of responses to online public opinion comment |
| | | Total number of online public opinion events forwarded |
| | | Total blog volume of online public opinion events |
| | | Number of blogs sent by the government |
| Information Environment | Topic Attention | Total reading of online public opinion events |
| | | Total discussion of online public opinion events |
| | Topic Activity | Change rate of online public opinion event blog public opinion events (*RVol*) |
| | | Change rate of online public opinion event blog forwarding (*RFor*) |
| | | Change rate of online public opinion event blog of Comment (*RRev*) |
| | | Change rate of online public opinion event blog of Like (*RLik*) |
| | | Change rate of online public opinion event blog of comment responses (*RRes*) |
| Information | Topic Sentiment Tendency | positive blog volume + Comment volume |
| | | Negative blog volume + comment volume |
| | | Neutral blog volume + comment volume |
| | | Change rate of online public opinion event blog of negative blog volume + comment (*RNeg*) |

$RVol = \frac{s_{t+\Delta t} - s_t}{\Delta t}$ is the change rate of the amount of blogs published about the same online public opinion event. Where $\Delta t$ is the time difference, $s_{t+\Delta t}$ is the total number of posts issued at $t + \Delta t$, $s_t$ is the total number of postings at $t$.

Since other indexes are similar to the above-mentioned indexes, here do not further explain them.

### 3.3. Design Date Collector

The data collector can collect online public opinion data automatically, and its working process is shown in Figure 4. In the first step, it accesses the page that includes the blog information of an online public opinion event for further obtaining data. In the second step, it downloads all page data, including blog content users information and so on. While in the third step, it extracts useful data from the downloaded pages and save it to the database. In the fourth step, it further processes and converts the acquired data, and then saves the processed data to their corresponding files.

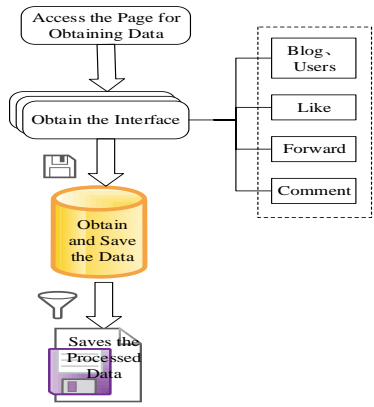

Figure 4. Data collection steps.

## 3.4. Index Quantification Method

In the initially constructed index system, the positive, negative and neutral indexes of topic emotion are not simple statistics [14, 17, 22]. In order to quantify these composite information, this paper adopts a sentiment classification model based on Long Short-Term Memory Network (LSTM) [15, 18], which has been shown in Figure 5.

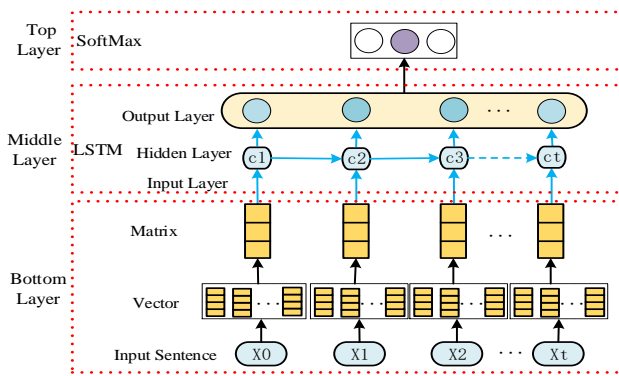

Figure 5. Word2Vec + LSTM sentiment classification model.

The bottom layer of the model uses word2vec to represent words with high-dimensional vectors, matrices to represent sentences, and to create word dictionaries that return the index of each word and the corresponding word index of each sentence.

The middle layer uses LSTM which can well contain sequence information to extract features. The middle layer is divided into three layers from top to bottom:

The first layer is the input layer of LSTM. At the bottom layer, words are represented by pre trained word vectors, sentences are represented by the matrix of word vectors, and the output sentence matrix is the input layer of this layer.

The second layer is the hidden layer, which uses the LSTM network structure to automatically extract features to make the semantic level of the features richer. One of the cruxes of the LSTM model is this "forgetting gate", which can control the convergence of the gradient during training (Thereby avoided the gradient vanishing exploding problem in RNN) while also maintaining long-term memory [1, 2, 16].

The third layer is the output layer of LSTM. The calculation process of LSTM is as follows:

$$
\begin{aligned}
f_t &= \sigma(W_f \cdot [S_{t-1}, x_t] + b_f) \\
i_t &= \delta(W_f \cdot [S_{t-1}, x_t] + b_i) \\
\widetilde{C}_t &= tanh(W_f \cdot [S_{t-1}, x_t] + b_c) \\
C_t &= f_t \circ C_{t-1}, x_t \circ C_t \\
O_t &= \sigma((W_f \cdot [S_{t-1}, x_t] + b_0) \\
S_t &= O_t \circ tanh(C_t)
\end{aligned}
\quad (1)
$$

Where $f_t$ is the information retained by LSTM after removing unnecessary information at time $t$, it is the information added by LSTM in t time stage. $\widetilde{c}_t$ is the newly added information updated by LSTM at time $t$. $c_t$ is the hidden state of the LSTM at time $t$, and the output of historical information at time $t$. $O_t$ is input at the $t$ time stage of LSTM. $S_t$ is output at the $t$ time stage of LSTM. $W_i$ and $b_i$ represent the cyclic weight matrix and offset matrix respectively. $\sigma(\cdot)$ is sigmoid activation function. $tanh(\cdot)$ is the activation function. The symbol $\circ$ indicate that the elements at the corresponding positions of the array are multiplied. The output $S_t$ of each time step of LSTM is not only related to the current input $o_t$ but also related to the hidden state $ct_{-1}$ the previous stage.

Hypothetical symbol $\theta$ represents all the parameters in the LSTM model, then for a given $x$ and $\theta$, the output layer transforms the result into a conditional probability distribution $P(y|x,\theta)$ for each element in the set y. Given the training set $T = \{(x(i), t(i), y(i)) \mid 1 \leq i \leq |T|\}$ and $y=\{y_1,y_2,y_3\}$. Suppose $y(i)$ is the model for the label value of the prediction result of the input $x(i)$, $t(i)$ is the real result, then $f(x(i),\theta)$ for each $yj\ (1 \leq j \leq 3)$ estimate their probability values $P(y_j|x(i),\theta)$ respectively, and output a normalized m-dimensional vector to represent the model's predicted probability distribution over the m label values:

$$
f(x^i, \theta) = \frac{1}{\sum_{j=1}^{3} \exp(P(y_i|x^i, \theta))} \begin{bmatrix} \exp(P(y_1 \mid x(i), \theta)) \\ \exp(P(y_2 \mid x(i), \theta)) \\ \exp(P(y_3 \mid x(i), \theta)) \end{bmatrix}
\quad (2)
$$

The top layer uses softmax for supervised classification training, as shown in formula (3).

$$
y^i = max(f(x(i), \theta))
\quad (3)
$$

## 3.5. Determine the Final Index System

There may be a problem of information superposition in the initially constructed index system, it is necessary to deduplicate the index in this paper. This article collects data on various typical online public opinion events given by experts. Through CA, the indexes with correlation coefficient exceeds the threshold in the same criterion layer are eliminated to reduce the duplication of screening results. And through the PCA, the indexes with small contribution are deleted to achieve the purpose that the selected indicators can have a greater impact on the evaluation results. The above method uses quantitative analysis instead of the expert qualitative analysis used in the general method, so the subjective one-sidedness of experts is avoided, and the selection result is relatively more objective.

### 3.5.1. Correlation Analysis Method

Since the data of the two indexes have a certain correlation, so we use Spearman correlation coefficient analysis to calculate the correlation between the two variables [2], and retain one of the two indexes with significant correlation, and one of the two indexes with significant correlation is retained. Spearman correlation coefficient (is called rank correlation coefficient), uses the rank size of two variables for correlation analysis. There is no requirement for the distribution of the original variables or for linearity, it can be calculated according to the following steps:

- *Step* 1: convert the elements corresponding to the two column vectors X and Y into $X_i$ and $Y_i$ into rankings in the respective column vector, denoted as $R(X_i)$ Sum $R(Y_i)$.
- *Step* 2: calculate the difference d between $R(X_i)$ and $R(Y_i)$ of the corresponding elements in the two column vectors $X$ and $Y$ according to formula (4), and add them.

$$d = \sum_{i=1}^{N} |R(X_i) - R(Y_j)|^2 \quad (4)$$

- *Step* 3: finally, calculate the correlation $R_s$ between the two column vectors according to formula (5).

$$R_S = 1 - \frac{6*d}{N*(N^2-1)} \quad (5)$$

### 3.5.2. Principal Component Analysis Method

PCA is a statistical method [4]. For all the variables originally proposed, delete the redundant variables (closely related variables) and create new variables as few as possible, so that these new variables are irrelevant, and these new variables reflect the information of the subject. Keep the original information as much as possible.

Calculation steps of PCA:

- *Step*1: Find the sample normalization matrix Z.

$$Z_{ij} = \frac{x_{ij} - \overline{x_j}}{s_j^2}, i = 1,2,...,n; j = 1,2,...,p \quad (6)$$

Where $\overline{x_j} = \frac{\sum_{i=1}^{n} x_{ij}}{n}$, $s_j^2 = \frac{\sum_{i=1}^{n}(x_{ij} - \overline{x_j})}{n-1}$, $x_{ij}$ is the index variable in the $i$-th row and $j$-column. $\overline{x_j}$ is the average number of indicators in column j. $s_j$ is the standard deviation of the index in column j. $Z_{ij}$ is the number of the $i$-th row and $j$-column in the standardized array Z.

- *Step* 2: Find the correlation coefficient matrix for the standardized matrix Z.

$$R = \frac{Z^T * Z}{n-1} \quad (7)$$

- *Step* 3: solve the characteristic equation of the sample correlation matrix $|R - \lambda I_p| = 0$, get $p$, $p$ is feature roots, and λ is the characteristic root of R. According to $\frac{\sum_{j=1}^{m} \lambda_j}{\sum_{j=1}^{p} \lambda_j}$, determine the number of principal components.

## 3.6. Building a Crisis Rating Model

Dividing the level of online public opinion crisis is the prerequisite for realizing the early warning of online public opinion crisis. The online public opinion crisis is divided into four levels according to the classification of the existing online public opinion early warning levels and the management regulations of relevant Chinese institutions [11, 25].

Giant police opinion (level 1): internet users pay close attention to online public opinion. Police opinion spread very quickly. It has become an event of online public opinion.

Serious police opinion (Level 2): internet users pay close attention to online public opinion. Police opinion's influence has spread to a wide range. Moreover, online public opinion is highly likely to be become an event of online public opinion.

Inter-mediate police opinion (Level 3): internet users have a high degree of attention to online public opinion. The speed of propagation is medium, and the influence of online public opinion is limited within a certain range. If there are no important nodes, they will not transform into online public opinion events.

Light police opinion (Level 4): Internet users have a low degree of attention to online public opinion. The impact is limited to a small range, and there is no possibility of turning it into behavioral opinion.

In order to quantify the crisis level, the benchmark values $X_i=(x_i(1), x_i(2), …, x_i(n))$, $(i=1,2,3,4)$, $(n=1, …, 13)$ were determined through the Delphi method. Where $x_i(n)$ is $n$-th indicator value of the benchmark of the $i$-th crisis level. The thirteen indicators are (likes, comments, reposts, posts, discussions, reads, microblogs, post change rate, repost change rate, likes change Rate, positive quantity, negative quantity, neutral quantity). $X_1$ =(4000, 3700, 3800, 500000, 1*109, 10000, 3.7, 4, 3.7, 4,700, 1800, 200). $X_2$=(3500, 3000, 3000, 30000, 1*107, 4000, 3.5, 3.2, 3, 3.3, 600, 1200, 250). $X_3$ =(3000, 2000, 1500, 15000, 1 *105, 2000, 3, 2.8, 2, 2.5, 800, 600, 280). $X_4$=(2000, 1000, 800, 10000, 1*104, 1000, 1.2, 1, 1, 1.5, 500, 200, 350). Then, the crisis level of the online public opinion event is determined by comparing the correlation between the online public opinion event data and the reference value. The specific work flow is as follows:

- *Step* 1, determine the reference value matrix:

$$(X_1, X_2, X_3, X_4) = \begin{pmatrix} x_1(1) & x_2(1) & x_3(1) & x_4(1) \\ x_1(2) & x_2(2) & x_3(2) & x_4(2) \\ \cdots & \cdots & \cdots & \cdots \\ x_1(n) & x_2(n) & x_3(n) & x_4(n) \end{pmatrix} \quad (8)$$

- *Step* 2: calculate the absolute value of the corresponding index in each evaluation target index data vector $x_0(k)$ and the reference value vector $x_i(k)$ of each index of each level. For the calculation method, see formula (9):

$$|x_0(k) - x_i(k)|, (k = 1,2,3, \ i = 1,2,\dots,n) \quad (9)$$

Where $x_0(k)$ is the k-th index data of the evaluation object; $x_i(k)$ is the benchmark value of the i-th level; i=1,2,3,4 are the four crisis levels, and k is the k-th index in the index system, N is the number of indexes in the index system.

- *Step* 3: calculate the minimum and maximum absolute values of each evaluation target index data and the reference value of the corresponding index of each level. For the calculation method, see formula (10):

$$\min_{x=1}^{n} |x_0(k) - x_i(k)| \text{ and } \max_{i=1}^{n} |x_0(k) - x_i(k)| \quad (10)$$

- *Step* 4: calculate the correlation coefficient $\xi_i(k)$ of each evaluation target index data and the corresponding reference value of each level index separately. For the calculation method, see formula (11):

$$\xi_i(k) = \frac{w_k \cdot \min_i |x_0(k) - x_i(k)| + \rho \cdot w_k \cdot \max_i |x_0(k) - x_i(k)|}{w_k \cdot |x_0(k) - x_i(k)| + \rho \cdot w_k \cdot \max_i |x_0(k) - x_i(k)|}, \quad (11)$$

k=1,2,…,n

Among them, $w_k$ is the weight of each data; ρ is the resolution coefficient, 0 <ρ <1.

- *Step* 5: calculate correlation coefficient. The calculation method is shown in formula (12):

$$\gamma_i = \frac{1}{n} \sum_{k=1}^{n} \xi_i(k) \quad (12)$$

## 4. Experiment

### 4.1. Quantification of Sentiment Index

The experimental sample is the Chinese text sentiment analysis data sets of Tan Songbo and Jia Jianbo downloaded from https://download.csdn.net/. There are 8033 positive texts, 8703 negative texts and 8355 neutral texts in the sample database, which divided into the test set and the training set of 8:2 to train the emotion classification model. The same time, in order to ensure the accuracy of the model, the cross-validation method is adopted, and eight of them are selected as the training set and the other two are used as the verification set in a free arrangement. The classification effect of the proposed method is shown in Table 2. Experimental result shows that the recall rate, accuracy rate and F1 values are above 85, which is higher than the traditional sentiment dictionary-based method, so the result of quantifying the sentiment index is relatively accurate.

Table 2. Sentiment classification effect of this method.

| model | precision | recall | F1 |
|---|---|---|---|
| Dictionary | 81.92% | 80.51% | 81.21% |
| Method of this paper | 86.61% | 88.00% | 87.30% |

### 4.2. Determine the Final Index System

This section analyzes the index based on the correlation analysis-principal component analysis method.

#### 4.2.1. Index Analysis Based on CA Method

Based on the data obtained from typical online public opinion events, using SPSS software, Spearman Rho correlation analysis method was used to screen the initial indexes, and the significant correlation threshold was set to 0.84 [24]. The correlation coefficients between the average number of ranks, the average number of fans, and the average number of historical blogs are 0.874 and 0.887, respectively, which are greater than 0.84, indicating that the two indexes have significant correlations, so the level index needs to be deleted, as shown in Table 3.

Table 3. CA of the internet user importance index.

|  | Avag Att | Avg Gra | Avg Foll | V num | AvgHis Vol |
|---|---|---|---|---|---|
| **Avag Att** | 1.000 | 0.763 | 0.729 | 0.560 | 0.740 |
| **AvgGra** |  | 1.000 | 0.874 | 0.713 | 0.887 |
| **AvgFol** |  |  | 1.000 | 0.819 | 0.812 |
| **V num** |  |  |  | 1.000 | 0.757 |
| **Avg His Vol** |  |  |  |  | 1.000 |

The correlation coefficients between the total number of responses to the index online public opinion event, the total number of blogs to the index online public opinion event, and the total number of comments to the online public opinion event are 0.873 and 0.935, which are greater than 0.84. Therefore, delete the index of the total number of responses to online public opinion events, as shown in Table 4.

Table 4. CA of the internet user participation index.

|  | NumBlog | Num For | NumLik | NumCom | NumRes | Num BG |
|---|---|---|---|---|---|---|
| **NumB** | 1.000 | 0.691 | 0.657 | 0.668 | 0.873 | 0.491 |
| **NumFor** |  | 1.000 | 0.833 | 0.816 | 0.796 | 0.598 |
| **NumLik** |  |  | 1.000 | 0.755 | 0.816 | 0.427 |
| **NumCom** |  |  |  | 1.000 | 0.935 | 0.721 |
| **NumRes** |  |  |  |  | 1.000 | 0.654 |
| **NumBG** |  |  |  |  |  | 1.000 |

The correlation coefficient between the change rate of the index online public opinion event comment reply and the change rate of the index online public opinion event comment is 0.891 greater than 0.84. Therefore, delete any of the two indexes. This paper deletes the index of the change rate of the total comment response of online public opinion events, as shown in Table 5.

Table 5. CA of the topic attention index.

|       | RVol  | RFor  | RRev  | RLik  | RRes  |
|-------|-------|-------|-------|-------|-------|
| RVol  | 1.000 | 0.733 | 0.651 | 0.782 | 0.883 |
| RFor  |       | 1.000 | .533  | 0.823 | 0.797 |
| RRev  |       |       | 1.000 | .731  | 0.891 |
| RLik  |       |       |       | 1.000 | 0.712 |
| RRes  |       |       |       |       | 1.000 |

The correlation coefficients between the remaining indexes are less than 0.84, as shown in Table 6 and Table 7, it shows that there is no problem of information overlap between indexes, and all indexes are retained.

Table 6. CA of the topic activity index.

|               | Num Reading | Num Discussion |
|---------------|-------------|----------------|
| NumReading    | 1.000       | 0.823          |
| NumDiscussion |             | 1.000          |

Table 7. CA of the topic sentiment tendency index.

|      | Neg   | Pos   | Neu   | RNeg    |
|------|-------|-------|-------|---------|
| Neg  | 1.000 | 0.533 | 0.351 | 0.782   |
| Pos  |       | 1.000 | 0.526 | -0.397[1] |
| Neu  |       |       | 1.000 | -0.151  |
| RNeg |       |       |       | 1.000   |

### 4.2.2. Index Analysis Based on PCA

According to the results of CA, use the PCA method to select the remaining index, this article sets the cumulative variance contribution rate threshold to 90%. Calculate the PCA results of Internet User Importance, as shown in Table 8, each index has a large contribution rate, so we need retain all indicators.

Table 8. PCA of the internet user importance index.

| Index   | Cumulative Contribution Rate | Contribution Rate | Characteristic Value |
|---------|------------------------------|-------------------|----------------------|
| AvgAtt  | 50.010                       | 50.010            | 4.311                |
| AvgFol  | 73.541                       | 23.531            | 2.097                |
| V num   | 87.354                       | 13.813            | 1.652                |
| AvgHisV | 100.000                      | 12.646            | 1.439                |

Calculate the contribution rate of the Internet User Participation Index, the contribution rate of the Topic Attention Index and the contribution rate of the Topic Activity Index, which can be seen in Table 9, Table 10 and Table 11, keep all index. Calculate the contribution rate of Topic Sentiment Tendency Index, the cumulative contribution rate of the first three PCA reaches 94.533, which are greater than 90%, as shown in Table 12, so retain the top three index with high contribution rate.

Table 9. PCA of the internet user participation index.

| Index  | Cumulative Contribution Rate | Contribution Rate | Characteristic Value |
|--------|------------------------------|-------------------|----------------------|
| NumB   | 40.727                       | 40.727            | 3.961                |
| NumFor | 61.105                       | 20.378            | 2.005                |
| NumLik | 76.216                       | 15.111            | 1.754                |
| NumCom | 89.521                       | 13.305            | 1.531                |
| NumBG  | 100.000                      | 10.476            | 0.927                |

Table 10. PCA of the topic attention index.

| Index | Cumulative Contribution Rate | Contribution Rate | Characteristic Value |
|-------|------------------------------|-------------------|----------------------|
| NumR  | 61.149                       | 61.149            | 4.853                |
| NumD  | 100.00                       | 35.851            | 3.147                |

Table 11. PCA of topic activity index.

| Index | Cumulative Contribution Rate | Contribution Rate | Characteristic Value |
|-------|------------------------------|-------------------|----------------------|
| RVol  | 50.487                       | 50.487            | 4.407                |
| RFor  | 75.856                       | 25.369            | 2.559                |
| RRev  | 88.673                       | 12.817            | 1.457                |
| RLik  | 100.000                      | 11.327            | 1.012                |

Table 12. PCA of the topic sentiment tendency index.

| Index | Cumulative Contribution Rate | Contribution Rate | Characteristic Value |
|-------|------------------------------|-------------------|----------------------|
| Neg   | 58.533                       | 58.533            | 4.782                |
| Pos   | 77.901                       | 19.368            | 1.917                |
| Neu   | 94.533                       | 16.632            | 1.781                |
| RNeg  | 100.000                      | 5.467             | 0.503                |

Firstly, after the CA method is used to eliminate the indexes that have significant correlation, the PCA method is used to concentrate the information. This process ensure that there is no superposition of information on the relevant indexes, verify the effectiveness of the index deletion method in the index system proposed in this paper. This paper build online public opinion evaluation index system, as shown in Table 13.

---

[1] Note: "-" is negatively correlated.

Table 13. The multi-level online public opinion evaluation index system constructed in this paper.

| First-level index | Secondary-level index | Third-level index |
|---|---|---|
| Information Person A1 | Internet User Importance B11 | Average followers (AvgFollower) C111 |
| | | Average attentions (AvgAttention) C112 |
| | | V number C113 |
| | | Average historical blog volume (AvgHisVol) C114 |
| | Internet User Participation B12 | Total number of likes of online public opinion events C121 |
| | | Total comments on online public opinion events C122 |
| | | Total number of online public opinion events forwarded C123 |
| | | Total blog volume of online public opinion events C124 |
| | | Number of blogs sent by the government C125 |
| Information Environment A2 | Topic Attention B21 | Total reading of online public opinion events C211 |
| | | Total discussion of public opinion events C212 |
| | Topic Activity B22 | Change rate of public opinion event blog online public opinion events (RVol) C221 |
| | | Change rate of online public opinion event blog forwarding (RFor) C222 |
| | | Change rate of online public opinion event blog of Comment (RRev) C223 |
| | | Change rate of online public opinion event blog of Like (RLik) C224 |
| Information A3 | Topic Sentiment Tendency B31 | positive blog volume + Comment volume C311 |
| | | Negative blog volume + comment volume C312 |
| | | Neutral blog volume + comment volume C313 |

## 4.3. Application of the Index System

Taking "Xi'an female Benz owner safeguards rights" event on Weibo (www.weibo.com) as an object, the effectiveness of the index system of this paper in the evaluation of online public opinion crisis is verified.

### 4.3.1. Sentiment Analysis of Internet Users

LSTM sentiment classification model is used to determine the sentiment polarity of online public opinion event blogs and comment texts. The polarities of online public opinion event blogs and comment texts in various periods are counted. Figure 6 is drawn to analyze the Internet user's sentiment tendency. Intuitively. Since the evening is rest time, there is hardly any development for this kind of civil disputes and online public opinion events. On the fourth day, online public opinion turned negative, it has become a huge negative online public opinion event at 12:00 on 4/14/2019.

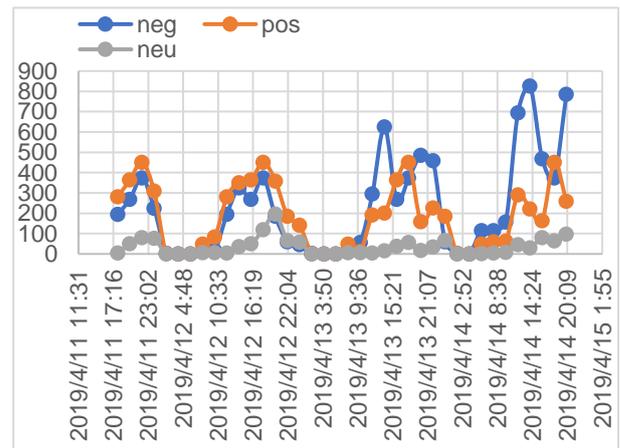

Figure 6. Sentiment changes trends of internet users.

### 4.3.2. Experimental Results and Analysis

From the second hour after the incident, according to the statistical data every two hours, Calculate the crisis level. The three indexes are C124, C211, and C212 dividing the crisis level. The seven indexes are C121, C122, C123, C124, C125, C211, C212 and the eleven indexes are respectively C121, C122, C123, C124, C125, C211, C212, C221, C222, C223, C224 and the fourteen indexes are C121, C122, C123, C124, C125, C211, C212, C221, C222, C223, C224, C311, C312, C313 and the eighteen indexes are C111, C112, C113, C114, C121, C122, C123, C124, C125, C211, C212, C221, C222, C223, C224, C311, C312 C313 dividing the crisis level. The monitoring results are shown in Figure 7. The four curves respectively represent the crisis level results under the four index systems. Eleven indexes add four rate of change indexes to seven indexes, the sensitivity of the classification has increased significantly. Fourteen indexes added three emotional indexes to eleven, which makes the ranking more sensitive. Eighteen indexes add four importance of Internet user indexes to fourteen indexes, to predict the level of the crisis approximately eight hours in advance, making the early warning effect better, and providing better help for crisis early warning. Verify that the index system of this paper is more sensitive and practical.

## 5. Discussion

In this paper, we constructed the multi-level online public opinion evaluation index system, which provides a theoretical basis for the online public opinion crisis early warning. The experimental results also confirmed the practicability of the evaluation method of the crisis level of online public opinion based on the index system. Nonetheless, the results of these experiments and the effectiveness of this method in other applications are also under the following threats:

1. Due to the fast updating speed of network vocabulary, the large number of new words will definitely reduce the accuracy of classification. Therefore, in order to ensure the effectiveness of the method in this paper, it is necessary to add training samples in time and expand the corpus.
2. The benchmark value and weight were determined by Delphi method. Therefore, the maxi-mum and minimum values of experts scoring are removed to minimize the impact.

What needs to be explained is that because the information ecosystem is a concept based on the theory of ecosystems, the analysis of the evolutionary situation of online public opinion and the analysis of the factors affecting the spread of online public opinion are based on the theory of information ecology. Weibo has the explosive power of diffusion.

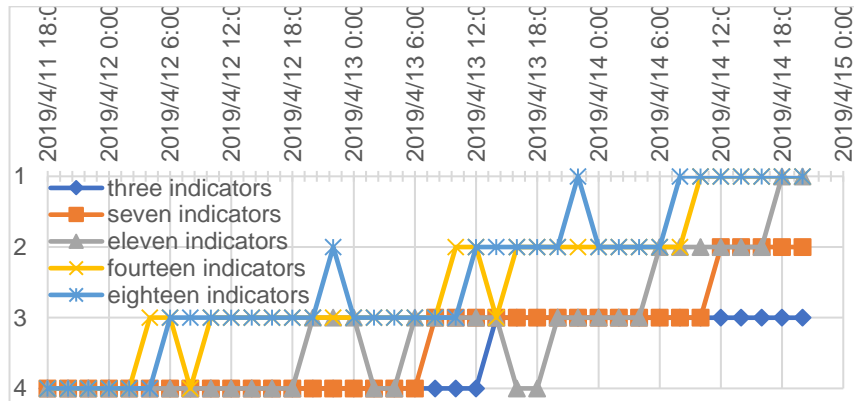

Figure 7. Online public opinion monitoring results.

## 6. Conclusions

This paper builds a multi-level online public opinion evaluation index system that includes online public opinion information, information person and information environment, which based on analyzing the influencing factors of online public opinion propagation from the perspective of information ecology. Compared with the existing index system, our index system is more comprehensive. In addition, all indexes have been screened by correlation and PCA so as to ensure that there is no information overlap in the entire indicator system. This index system can provide a comprehensive and objective evaluation standard for assessing the crisis level of online public opinion.

This paper builds a sentiment classification model of online public opinion text based on LSTM that makes the sentiment index in the index system quantifiable. In addition, the index system contains not only simple counting information, but also change rate information. On this basis, this paper divides the crisis level of online public opinion into 4 levels, and determines the benchmark value of each indicator through the Delphi method, which lays the foundation for early crisis warning. The experimental results also prove that the method in this paper can objectively evaluate the crisis level of online public opinion.

## Acknowledgement

This paper is one of the research results of Jilin Science and Technology Development Plan Project Network Public Opinion Analysis and Dynamic Evolution Mechanism Research for Public Crisis Early Warning (20190303107SF).

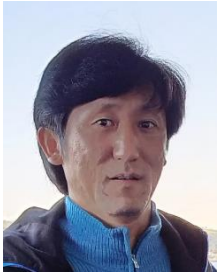

**Fanqi Meng** is an associate professor at NEEPU. He was born in Tongliao, Inner Mongolia, China in 1981. He received his Ph.D. degree in computer application technology from Harbin Institute of Technology, Harbin, in 2018. His research interests include software safety, natural language processing, fault diagnosis of electric power equipment and other aspects, involve software engineering, artificial intelligence, data mining and other fields.

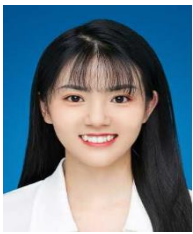

**Xixi Xiao** was born in Shangqu, Henan, China in 1995. She received the M.E. degree from Northeast Electric Power University. Her research interest include online public opinion and artificial intelligence.

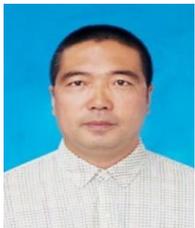

**Jingdong Wang** is an associate professor at NEEPU. He was born in Changchun, Jilin, China in 1980. He received the Ph.D. degree in information science from University of Science and technology of China, in 2017. His research interests include public security, natural language processing, knowledge graph and other aspects.